\documentclass[onecolumn,showpacs]{revtex4}

\usepackage{graphicx}
\usepackage{dcolumn}
\usepackage{bm}

\input epsf

\begin{document}

\title{\Large Tachyonic field interacting with Scalar (Phantom) Field}

\author{\bf  Surajit
Chattopadhyay$^1$\footnote{surajit$_{_{-}}2008$@yahoo.co.in} and
Ujjal Debnath$^2$\footnote{ujjaldebnath@yahoo.com ,
ujjal@iucaa.ernet.in}}

\affiliation{$^1$Department of Computer Application, Pailan
College of Management and Technology, Bengal Pailan Park,
Kolkata-700 104, India.\\
$^2$Department of Mathematics, Bengal Engineering and Science
University, Shibpur, Howrah-711 103, India. }

\date{\today}

\begin{abstract}
In this letter, we have considered the universe is filled with the
mixture of tachyonic field and scalar or phantom field. If the
tachyonic field interacts with scalar or phantom field, the
interaction term decays with time and the energy for scalar field
is transferred to tachyonic field or the energy for phantom field
is transferred to tachyonic field. The tachyonic field and scalar
field potentials always decrease, but phantom field potential
always increases.
\end{abstract}

\pacs{}

\maketitle

Recent measurements of the luminosity-redshift relations observed
[1, 2] for a number of newly discovered type Ia supernova indicate
that at present the universe is expanding in a accelerated
manner. This has given rise to a lot of dark energy models [3-6],
which are supposed to be the reason behind this present
acceleration. This mysterious fluid called dark energy is
believed to dominate over the matter content of the Universe by
70 $\%$ and to have enough negative pressure as to drive present
day acceleration. Most of the dark energy models involve one or
more scalar fields with various actions and with or without a
scalar field potential [7]. The ratio $w$ between the pressure
and the energy density of the dark energy seems to be near of
less than $-1$, $-1.62<w<-0.72$ [8]. Numerous models of dark
energy exist. There is much interest now in the tachyon cosmology
[9] where the appearance of tachyon is basically motivated by
string theory [10]. It has been recently shown by Sen [11, 12]
that the decay of an unstable D-brane produces pressure-less gas
with finite energy density that resembles classical dust. The
cosmological effects of the tachyon rolling down to its ground
state have been discussed by Gibbons [13]. Rolling tachyon matter
associated with unstable D-branes has an interesting equation of
state which smoothly interpolates between $-1$ and 0 i.e.,
$-1<w<0$. As the Tachyon field rolls down the hill, the universe
experiences accelerated expansion and at a particular epoch the
scale factor passes through the point of inflection marking the
end of inflation [10]. The tachyonic matter might provide an
explanation for inflation at the early epochs and could
contribute to some new form of cosmological dark matter at late
times [14]. Inflation under tachyonic field has also been
discussed in ref. [9, 15, 16]. Also the tachyon field has a
potential which has an unstable maximum at the origin and decays
to almost zero as the field goes to infinity. Depending on
various forms of this potential following this asymptotic
behaviour a lot of works have been carried out on tachyonic dark
energy [6, 17]. Sami et al [18] have discussed the cosmological
prospects of rolling tachyon with exponential potential.\\

The phantom field (with negative kinetic energy) [19] was also
proposed as a candidate for dark energy as it admits sufficient
negative pressure ($w<-1$). One remarkable feature of the phantom
model is that the universe will end with a ``big rip" (future
singularity). That is, for phantom dominated universe, its total
lifetime is finite. Before the death of the universe, the phantom
dark energy will rip apart all bound structures like the Milky
Way, solar system, Earth and ultimately the molecules, atoms,
nuclei and nucleons of which we are composed.\\

To obtain a suitable evolution of the Universe an interaction is
often assumed such that the decay rate should be proportional to
the present value of the Hubble parameter for good fit to the
expansion history of the Universe as determined by the Supernovae
and CMB data [20]. These kind of models describe an energy flow
between the components so that no components are conserved
separately. There are several work on the interaction between dark
energy (tachyon or phantom) and dark matter [21], where
phenomenologically introduced different forms of interaction term.\\

Here, we consider a model which comprises of a two component
mixture. Here we are interested in how such an interaction
between the tachyon and scalar or phantom dark energy affects the
evolution and total lifetime of the universe. We consider an
energy flow between them by introducing an interaction term which
is proportional to the product of the Hubble parameter and the
density of the tachyonic field.\\

The metric of a spatially flat isotropic and homogeneous Universe
in FRW model is

\begin{equation}
ds^{2}=dt^{2}-a^{2}(t)\left[dr^{2}+r^{2}(d\theta^{2}+sin^{2}\theta
d\phi^{2})\right]
\end{equation}

where $a(t)$ is the scale factor.\\

The Einstein field equations are (choosing $8\pi G=c=1$)

\begin{equation}
3H^{2}=\rho_{tot}
\end{equation}
and
\begin{equation}
6(\dot{H}+H^{2})=-(\rho_{tot}+3p_{tot})
\end{equation}

where, $\rho_{tot}$ and $p_{tot}$ are the total energy density and
the pressure of the Universe and $H=\frac{\dot{a}}{a}$ is the Hubble parameter.\\

The energy conservation equation is
\begin{equation}
\dot{\rho}_{tot}+3H(\rho_{tot}+p_{tot})=0
\end{equation}

Now we consider a two fluid model consisting of tachyonic field
and scalar field (or phantom field). Hence the total energy
density and pressure are respectively given by
\begin{equation}
\rho_{tot}=\rho_{1}+\rho_{2}
\end {equation}
and
\begin{equation}
p_{tot}=p_{1}+p_{2}
\end {equation}

The energy density $\rho_{1}$ and pressure $p_{1}$ for tachyonic
field $\phi_{1}$ with potential $V_{1}(\phi_{1})$ are respectively
given by

\begin{equation}
\rho_{1}=\frac{V_{1}(\phi_{1})}{\sqrt{1-{\dot{\phi}_{1}}^{2}}}
\end{equation}
and
\begin{equation}
p_{1}=-V_{1}(\phi_{1}) \sqrt{1-{\dot{\phi}_{1}}^{2}}
\end{equation}

The energy density $\rho_{2}$ and pressure $p_{2}$ for scalar
field (or phantom field) $\phi_{2}$ with potential
$V_{2}(\phi_{2})$ are respectively given by

\begin{equation}
\rho_{2}=\frac{\epsilon}{2}~\dot{\phi}_{2}^{2}+V_{2}(\phi_{2})
\end{equation}
and
\begin{equation}
p_{2}=\frac{\epsilon}{2}~\dot{\phi}_{2}^{2}-V_{2}(\phi_{2})
\end{equation}

where, $\epsilon=1$ for scalar field and $\epsilon=-1$ for
phantom field.\\

Therefore, the conservation equation reduces to
\begin{equation}
\dot{\rho}_{1}+3H(\rho_{1}+p_{1})=-Q
\end{equation}
and
\begin{equation}
\dot{\rho}_{2}+3H(\rho_{2}+p_{2})=Q
\end{equation}

where, $Q$ is the interaction term. For getting convenience while
integrating equation (11), we have chosen
$Q=3\delta H\rho_{1}$ where $\delta$ is the interaction parameter.\\

Now eq.(11) reduces to the form

\begin{equation}
\frac{\dot{V}_{1}}{V_{1}}+\frac{\dot{\phi}_{1}\ddot{\phi}_{1}}
{1-\dot{\phi}_{1}^{2}}+3H(\delta+\dot{\phi}_{1}^{2})=0
\end{equation}

Here $V_{1}$ is a function of $\phi_{1}$ which is a function of
time $t$. Naturally, $\dot{\phi}_{1}$ will be a function of time
$t$ and hence it is possible to choose $V_{1}$ as a function of
$\dot{\phi}_{1}$. Now, in order to solve the equation (13), we
take a simple form of
$V_{1}=\left(1-\dot{\phi}_{1}^{2}\right)^{-m}, ~(m>0)$ [22], so
that the solution of $\dot{\phi}_{1}$ becomes

\begin{equation}
\dot{\phi}_{1}^{2}=\left[-\delta+\left(\frac{c}{a^{3}}
\right)^{\frac{2(1+\delta)}{1+2m}}\right]\left[1+\left(\frac{c}{a^{3}}
\right)^{\frac{2(1+\delta)}{1+2m}}\right]^{-1}
\end{equation}
where $c$ is an integration constant. The potential $V_{1}$ of the
tachyonic field $\phi_{1}$ can be written as

\begin{equation}
V_{1}=\left[1+\left(\frac{c}{a^{3}}
\right)^{\frac{2(1+\delta)}{1+2m}} \right]^{m}(1+\delta)^{-m}
\end{equation}

So from equations (2) and (3), we have

\begin{equation}
\dot{\phi}_{2}^{2}=-\frac{2\dot{H}}{\epsilon}+\frac{1}{\epsilon}\left[1+\left(\frac{c}{a^{3}}
\right)^{\frac{2(1+\delta)}{1+2m}}\right]^{m-\frac{1}{2}}
\left[\delta-\left(\frac{c}{a^{3}}
\right)^{\frac{2(1+\delta)}{1+2m}}\right](1+\delta)^{-m-\frac{1}{2}}
\end{equation}
and
\begin{equation}
V_{2}=\dot{H}+3H^{2}-\frac{1}{2}\left[1+\left(\frac{c}{a^{3}}
\right)^{\frac{2(1+\delta)}{1+2m}}\right]^{m-\frac{1}{2}}
\left[2+\delta+\left(\frac{c}{a^{3}}
\right)^{\frac{2(1+\delta)}{1+2m}}\right](1+\delta)^{-m-\frac{1}{2}}
\end{equation}

Now eq.(12) can be re-written as

\begin{equation}
\dot{V}_{2}+\epsilon\dot{\phi}_{2}\ddot{\phi}_{2}+3H(\epsilon\dot{\phi}_{2}^{2}-\delta\rho_{1})=0
\end{equation}

Now putting the values of $\dot{\phi}_{2}$ and $V_{2}$, the
eq.(18) is automatically satisfied.\\

Now for simplicity, let us consider $V_{2}=n\dot{\phi}_{2}^{2}$,
so from the above equation (17) we have

$$
\dot{\phi}_{2}^{2}=c_{1}^{2}a^{-\frac{6\epsilon}{2n+\epsilon}}+\frac{6\delta}{2n+\epsilon}
(1+\delta)^{-m-\frac{1}{2}}a^{-\frac{6\epsilon}{2n+\epsilon}} \int
a^{\frac{6\epsilon}{2n+\epsilon}-1} \left[1+\left(\frac{c}{a^{3}}
\right)^{\frac{2(1+\delta)}{1+2m}}\right]^{m+\frac{1}{2}}da~~~~~~~~~~~~~~~~~~~~~~~~
$$

\begin{equation}
=c_{1}^{2}a^{-\frac{6\epsilon}{2n+\epsilon}}-\delta
(1+\delta)^{-m-\frac{1}{2}}~_{2}F_{1}[\frac{1+2m}{(1+\delta)(2n+\epsilon)},
-m-\frac{1}{2}, 1+\frac{1+2m}{(1+\delta)(2n+\epsilon)},
-\left(\frac{c}{a^{3}} \right)^{\frac{2(1+\delta)}{1+2m}} ]
\end{equation}
and
\begin{equation}
V_{2}=nc_{1}^{2}a^{-\frac{6\epsilon}{2n+\epsilon}}-n\delta
(1+\delta)^{-m-\frac{1}{2}}~_{2}F_{1}[\frac{1+2m}{(1+\delta)(2n+\epsilon)},
-m-\frac{1}{2}, 1+\frac{1+2m}{(1+\delta)(2n+\epsilon)},
-\left(\frac{c}{a^{3}} \right)^{\frac{2(1+\delta)}{1+2m}} ]
\end{equation}

\begin{figure}
\includegraphics[height=1.8in]{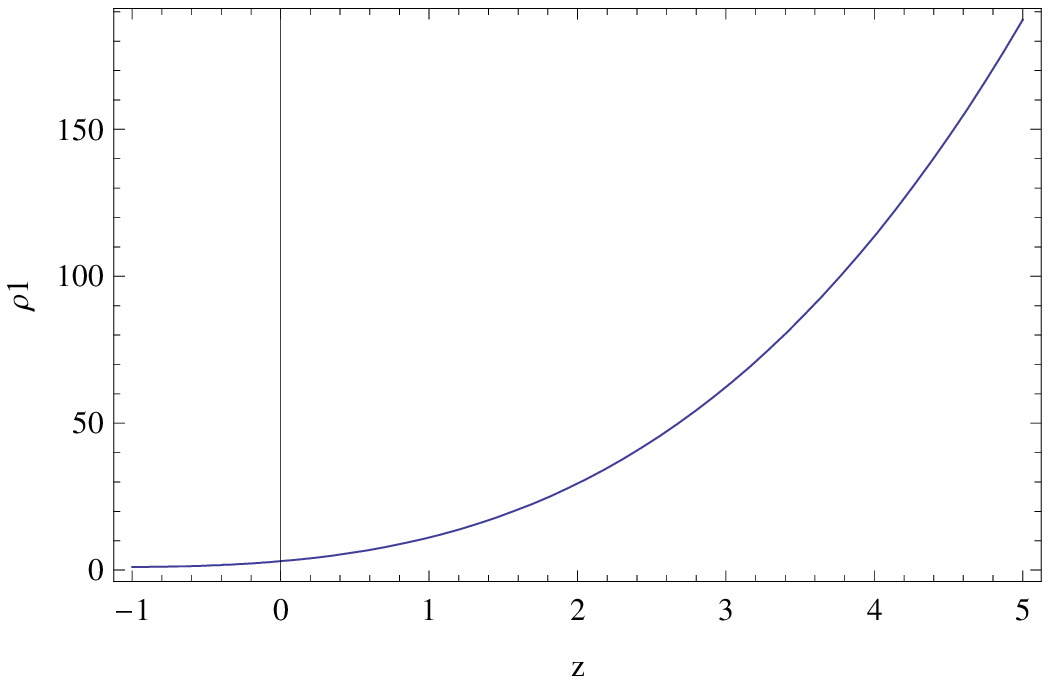}~~~~
\includegraphics[height=1.8in]{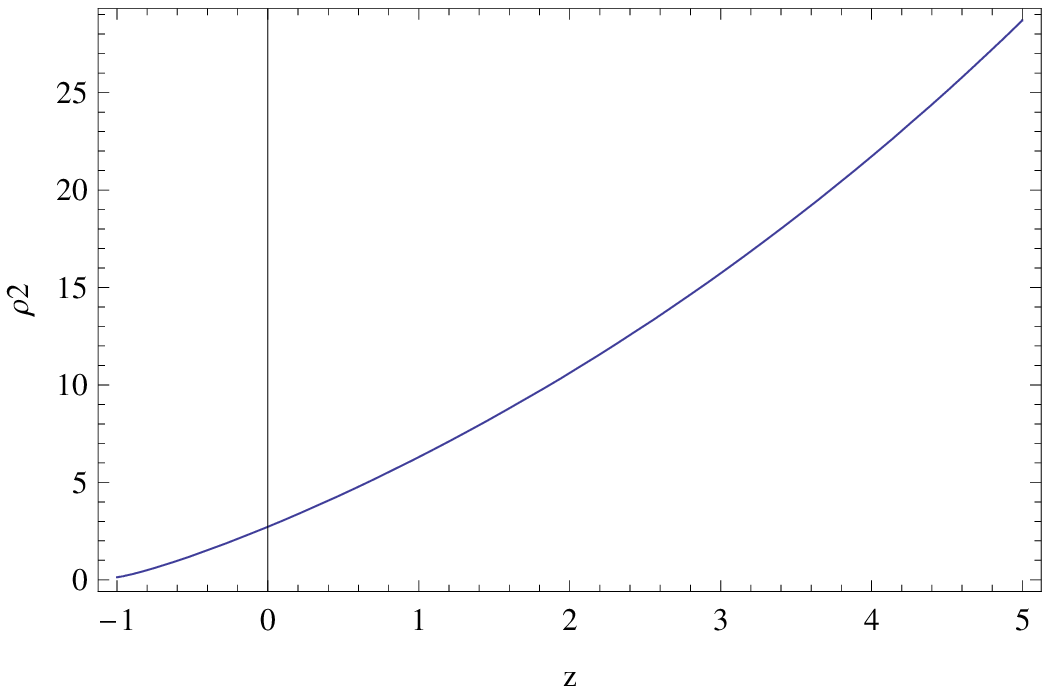}\\
\vspace{1mm} ~~~~~~~~~~~~Fig.1~~~~~~~~~~~~~~~~~~~~~~~~~~~~~~~~~~~~~~~~~~~~~~~~~~~~~~~~~~~~~~~~~~~~Fig.2\\

\vspace{7mm}

\includegraphics[height=1.8in]{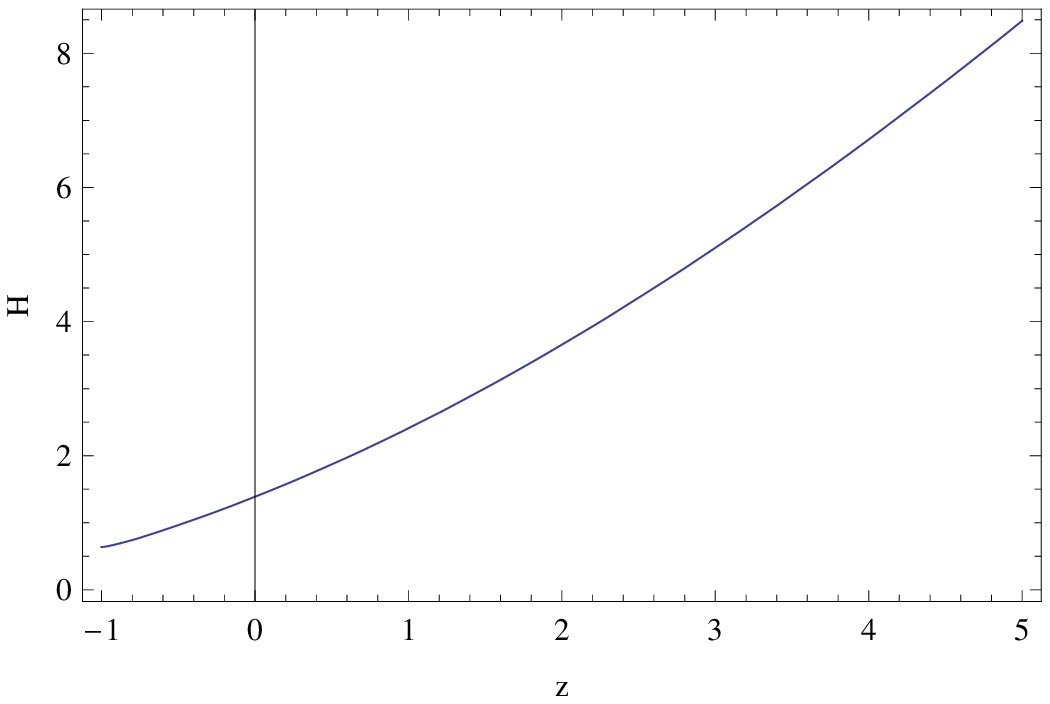}~~~~
\includegraphics[height=1.8in]{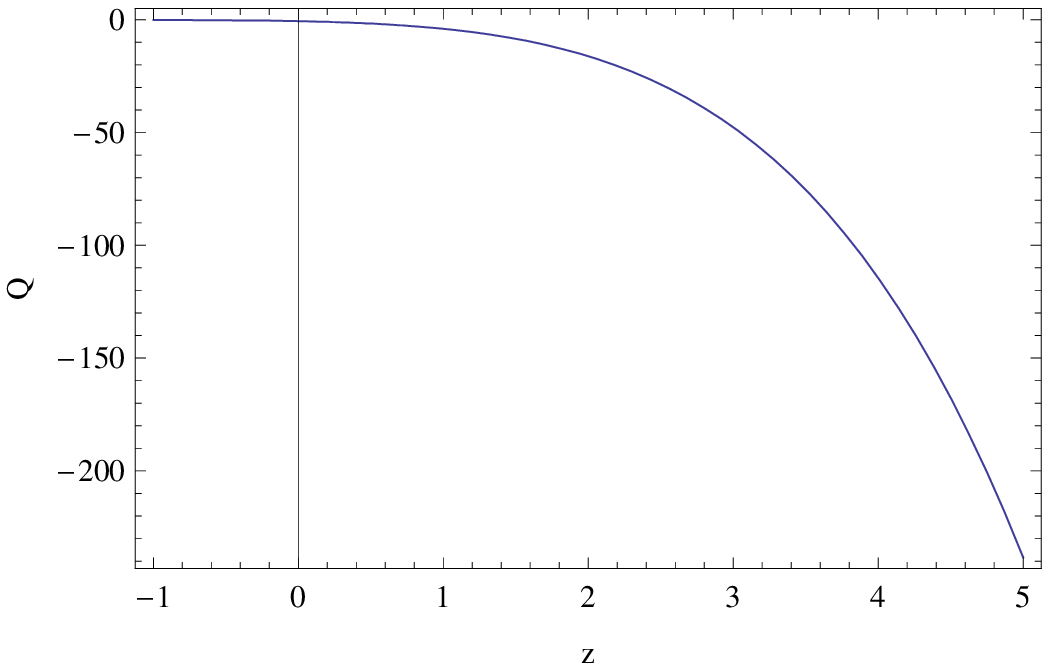}\\
\vspace{1mm} ~~~~~~~~~~~~~Fig.3~~~~~~~~~~~~~~~~~~~~~~~~~~~~~~~~~~~~~~~~~~~~~~~~~~~~~~~~~~~~~~~~~~~~~~~~~Fig.4\\

\vspace{7mm}

\includegraphics[height=1.8in]{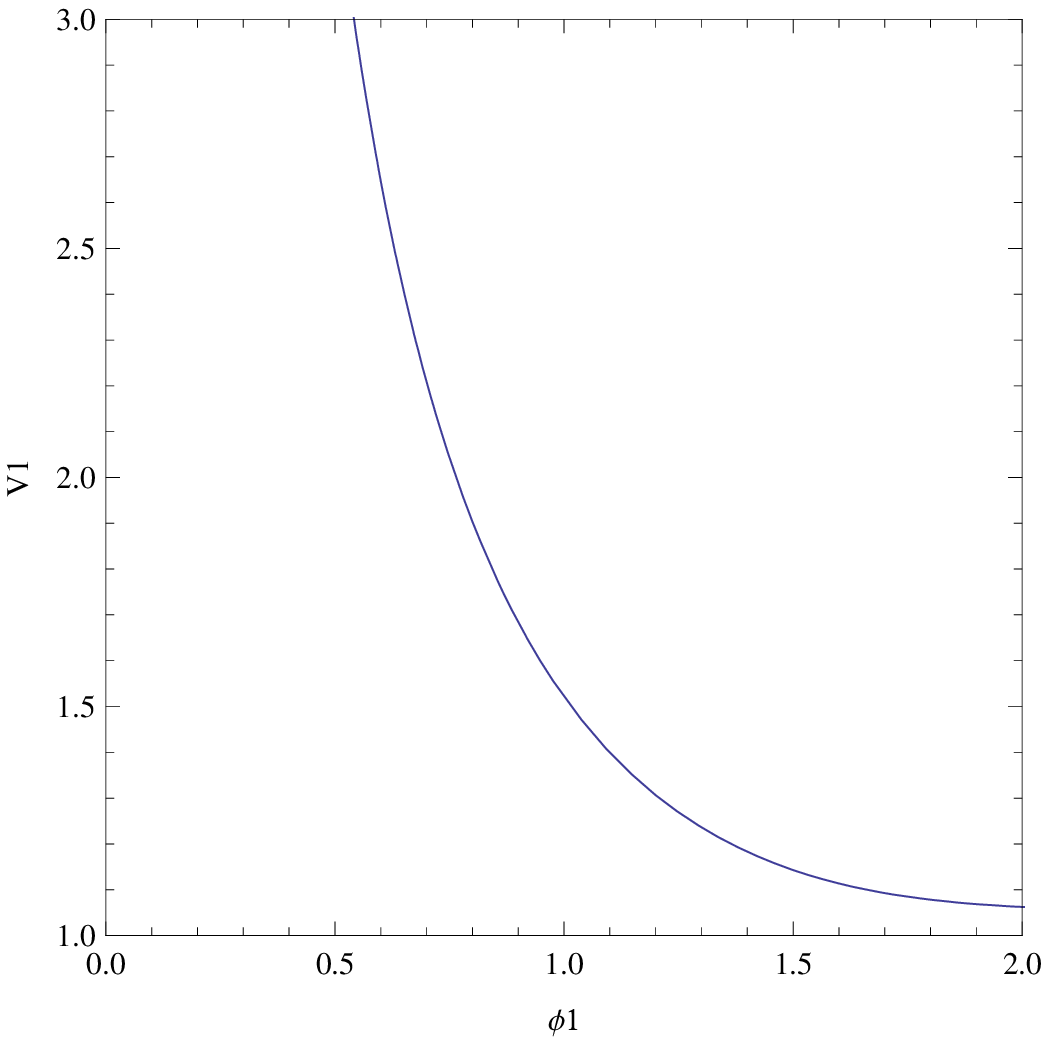}~~~~~~~~~~~~~~~~~~~
\includegraphics[height=1.8in]{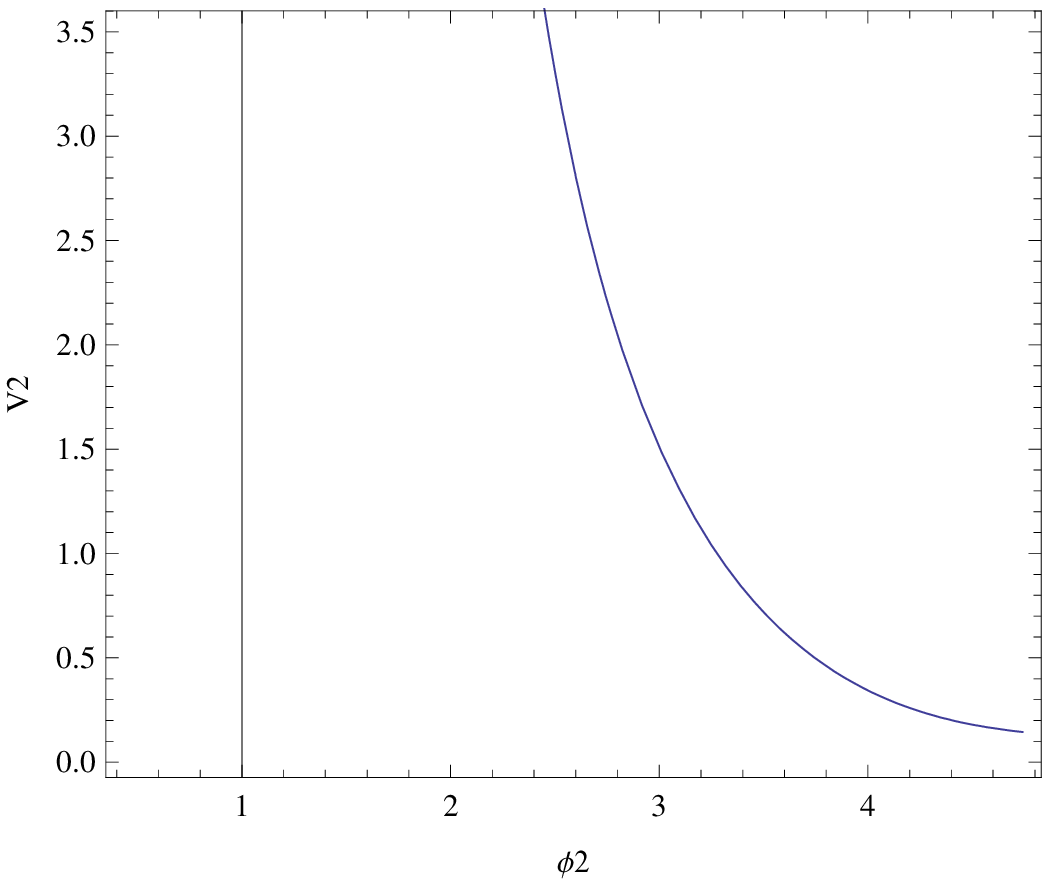}\\
\vspace{1mm} ~~~~~~~~~~~~~Fig.5~~~~~~~~~~~~~~~~~~~~~~~~~~~~~~~~~~~~~~~~~~~~~~~~~~~~~~~~~~~~~~~~Fig.6\\

\vspace{7mm}

Figs. 1, 2, 3 and 4 show variations of $\rho_{1},~\rho_{2},~H$ and
$q$ respectively with redshift $z$ and figs. 5, 6 show variations
of $V_{1}$ with $\phi_{1}$, $V_{2}$ with $\phi_{2}$ respectively
for
$\epsilon=+1, n=2, \delta=-0.05, m=c=c1=1$.\\

\vspace{6mm}

\end{figure}

Figures 1 - 6 are drawn for scalar field model with $\delta=-0.05$
and figures 7 - 12 are drawn for phantom model with
$\delta=-0.05$. Figs. 1 - 4 show the variations
$\rho_{1},~\rho_{2},~H,~Q$ with redshift $z$ and figs. 5, 6 show
the variations of $V_{1}$ with $\phi_{1}$, $V_{2}$ with $\phi_{2}$
respectively for scalar field model. Figs. 7 - 10 show the
variations $\rho_{1},~\rho_{2},~H,~Q$ with redshift $z$ and figs.
11, 12 show the variations of $V_{1}$ with $\phi_{1}$, $V_{2}$
with $\phi_{2}$ respectively for phantom model. For scalar field
model $\rho_{1},~\rho_{2},~H,~Q$ decrease with decreasing $z$ but
for phantom model $\rho_{1},~\rho_{2},~H$ decrease first and then
increase with decreasing $z$ and $Q$ decreases with decreasing
$z$. The tachyonic field and scalar field potentials always
decrease, but phantom field potential always increases.  For
scalar field model ($\epsilon=+1$) and phantom phantom field model
($\epsilon=-1$), $\delta$ may be negative due to positivity of
$\dot{\phi}_{2}^{2}$. So from equations (11) and (12), it may be
concluded that the energy for scalar field will be transferred to
tachyonic field the energy for phantom field will be transferred
to tachyonic field. In both the cases, the interaction term is
decays with time. Figures 13 - 16 are generated taking $\delta=0$.
Comparing figures 5, 6, 11 and 12 with figures 13, 14, 15 and 16,
we have seen that interaction does not have any significant impact
on the variations of potentials against fields. This may be due to the fact that
interaction is very small.\\

\begin{figure}
\includegraphics[height=1.8in]{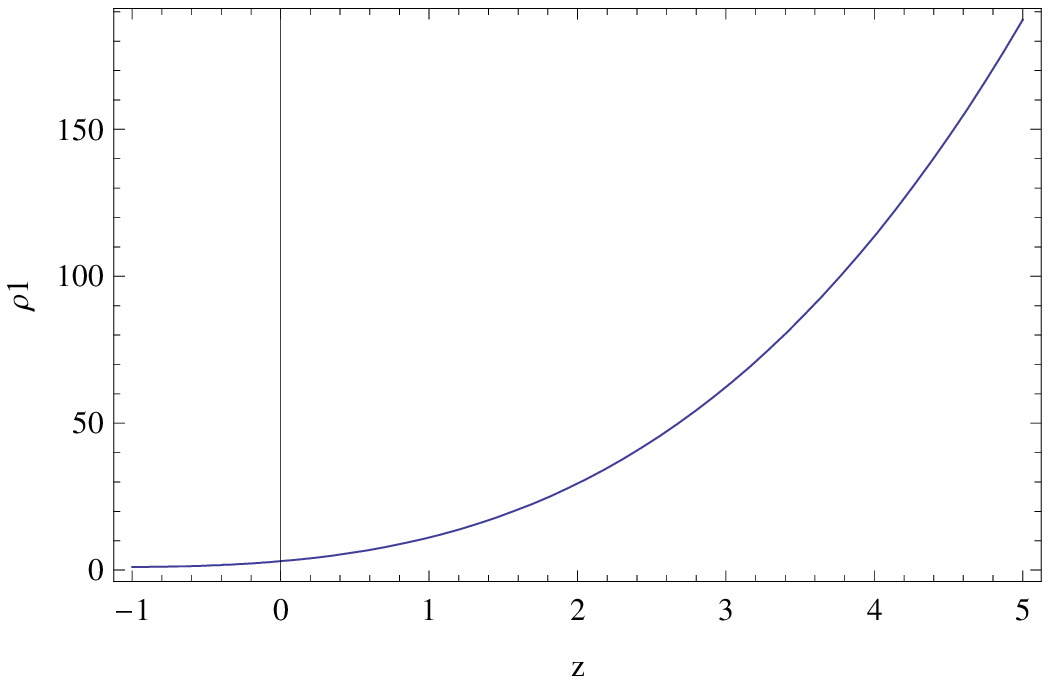}~~~~
\includegraphics[height=1.8in]{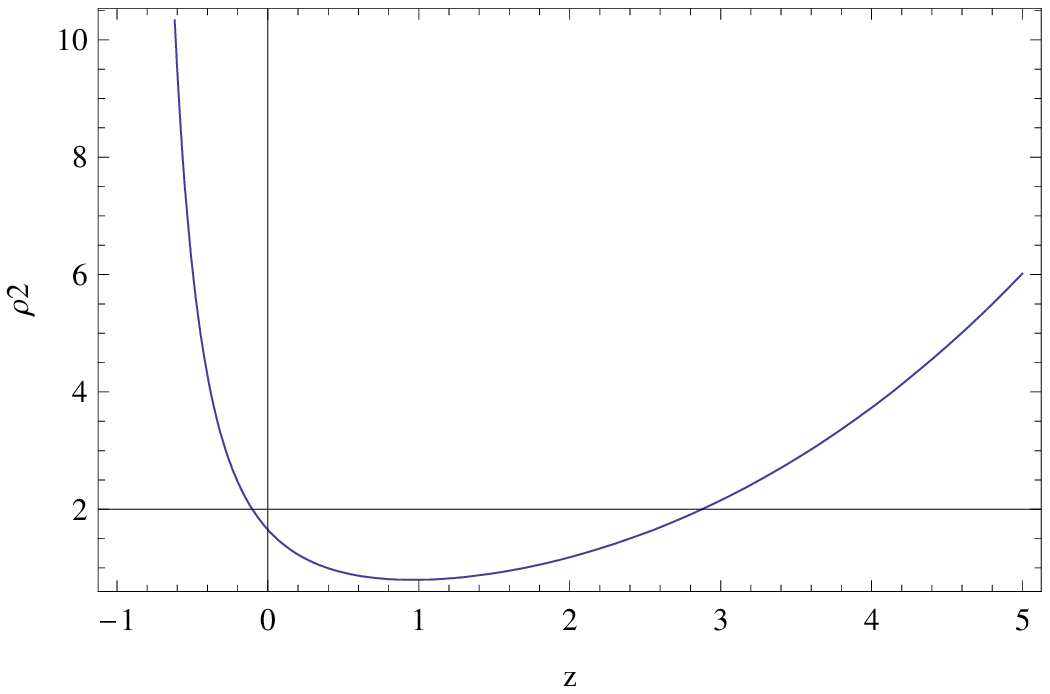}\\
\vspace{1mm} ~~~~~~~~~~~~Fig.7~~~~~~~~~~~~~~~~~~~~~~~~~~~~~~~~~~~~~~~~~~~~~~~~~~~~~~~~~~~~~~~~~~~~~~~Fig.8\\

\vspace{7mm}

\includegraphics[height=1.8in]{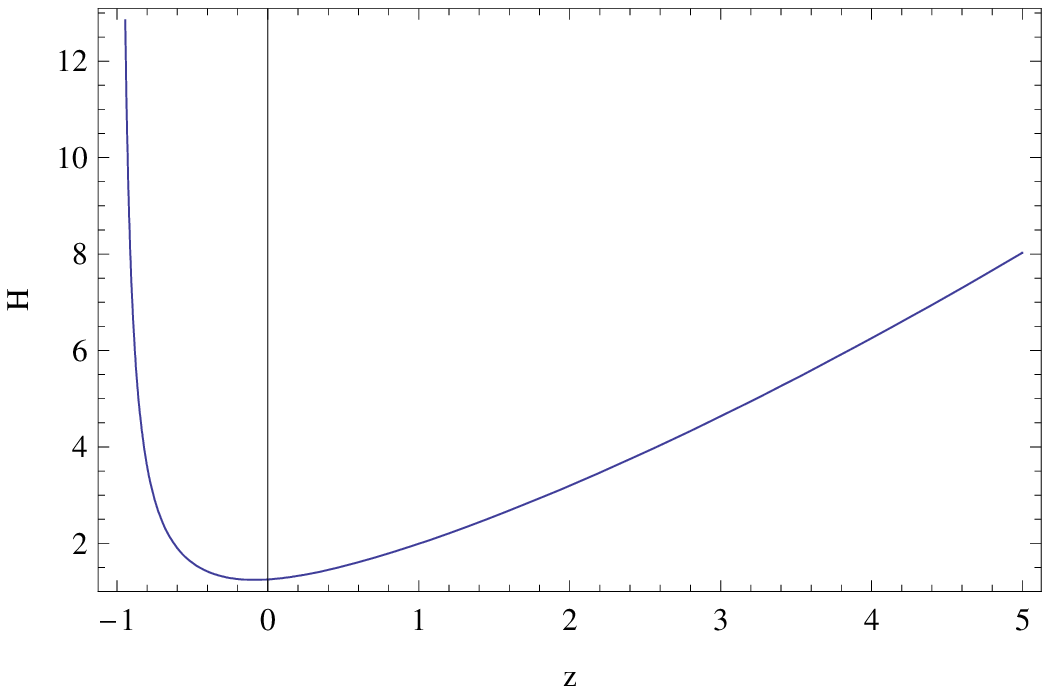}~~~~
\includegraphics[height=1.8in]{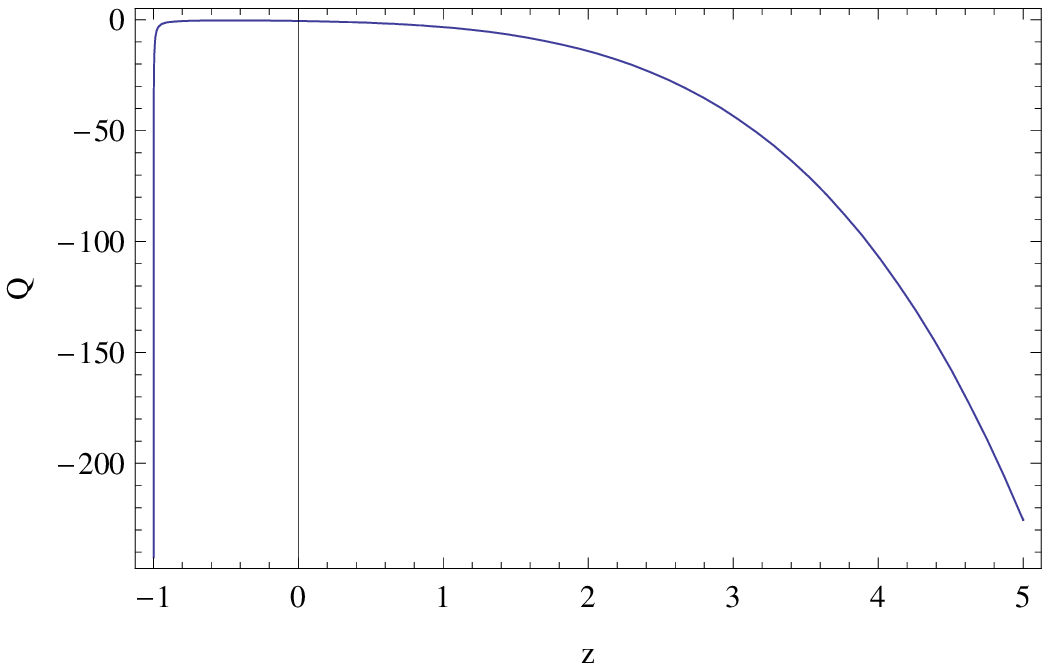}\\
\vspace{1mm} ~~~~~~~~~~~~~Fig.9~~~~~~~~~~~~~~~~~~~~~~~~~~~~~~~~~~~~~~~~~~~~~~~~~~~~~~~~~~~~~~~~~~~~~~Fig.10\\

\vspace{7mm}

\includegraphics[height=1.8in]{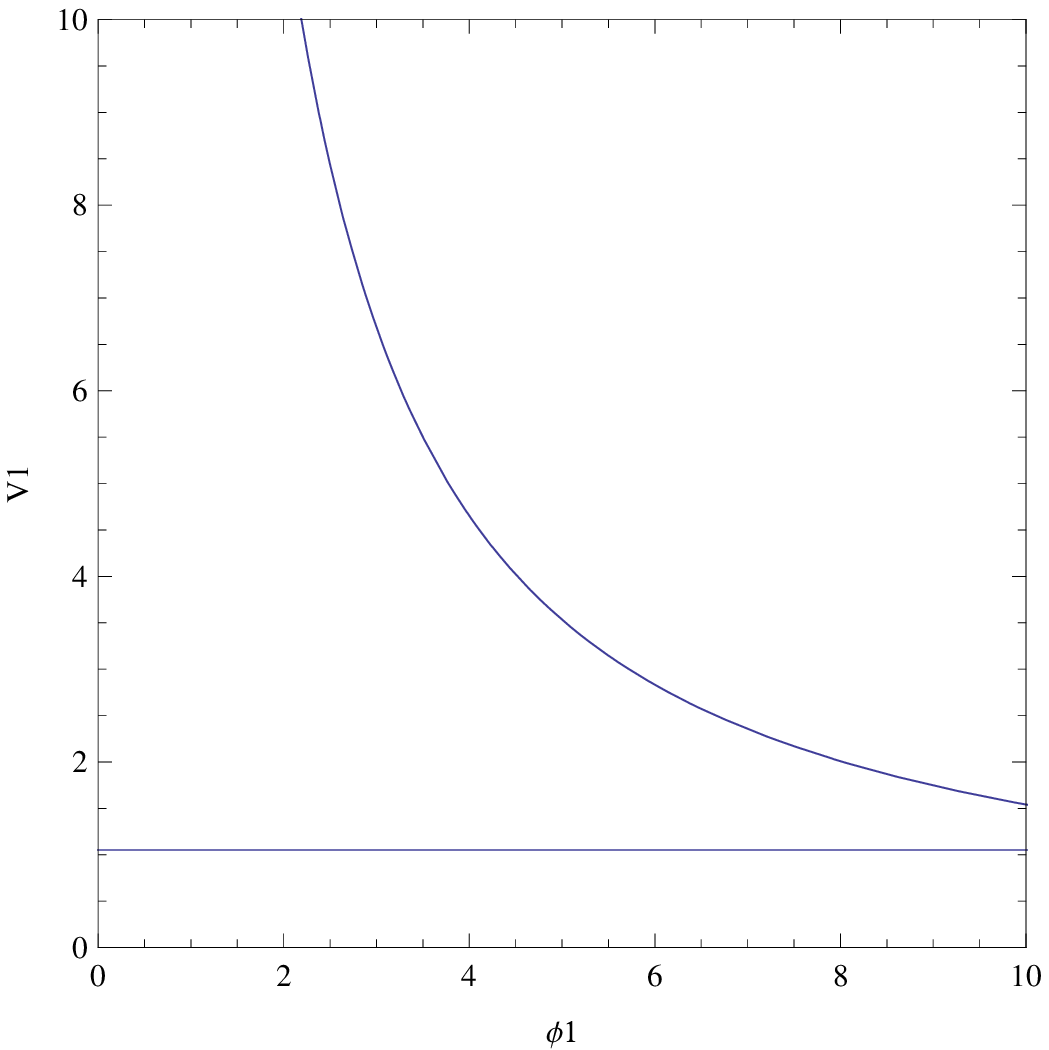}~~~~
\includegraphics[height=1.8in]{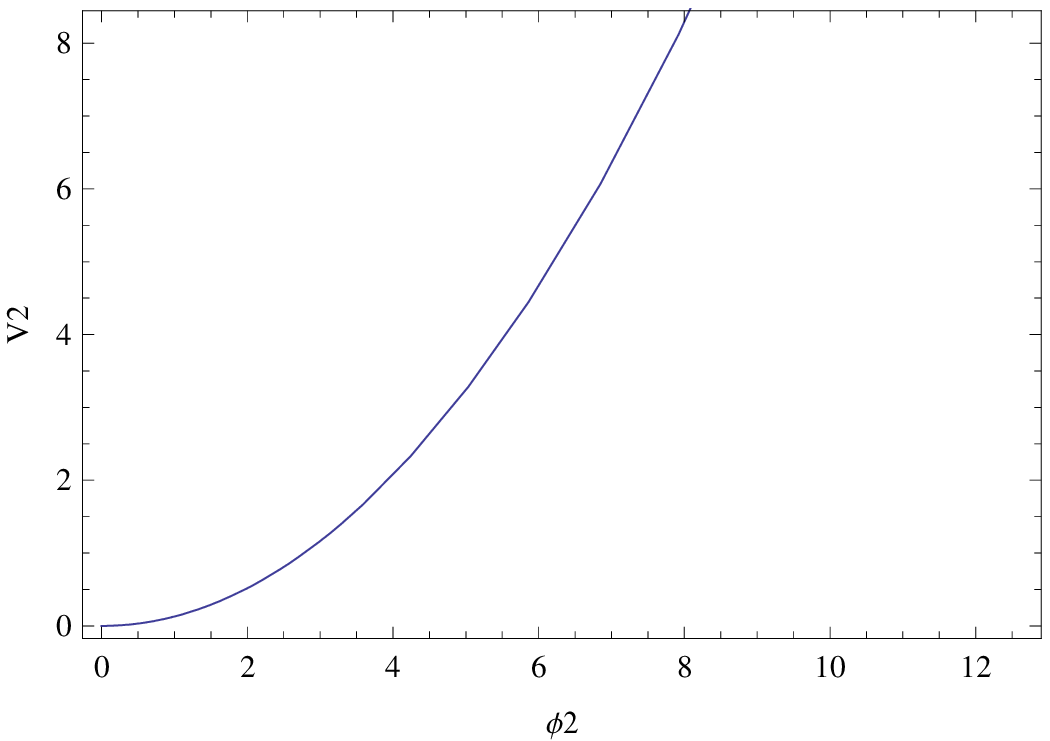}\\
\vspace{1mm} ~~~~~~~~~~~~~~~~~~Fig.11~~~~~~~~~~~~~~~~~~~~~~~~~~~~~~~~~~~~~~~~~~~~~~~~~~~~~~~~~~~~~~~~~~~~~~Fig.12\\

\vspace{7mm}

Figs. 7, 8, 9 and 10 show variations of $\rho_{1},~\rho_{2},~H$
and $q$ respectively with redshift $z$ and figs. 11, 12 show
variations of $V_{1}$ with $\phi_{1}$, $V_{2}$ with $\phi_{2}$
respectively for
$\epsilon=-1, n=2, \delta=-0.05, m=c=c1=1$.\\

\vspace{6mm}

\end{figure}

\begin{figure}
\includegraphics[height=1.8in]{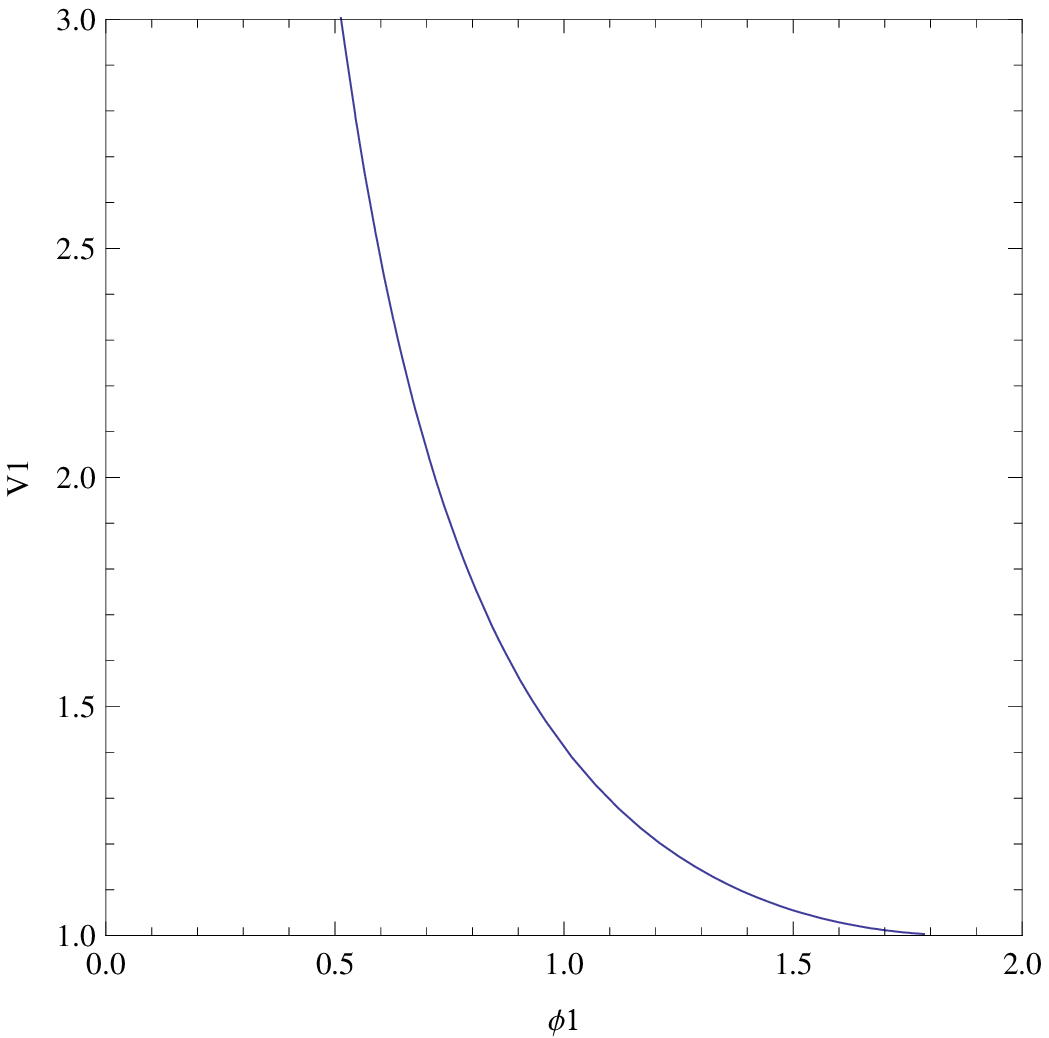}~~~~
\includegraphics[height=1.8in]{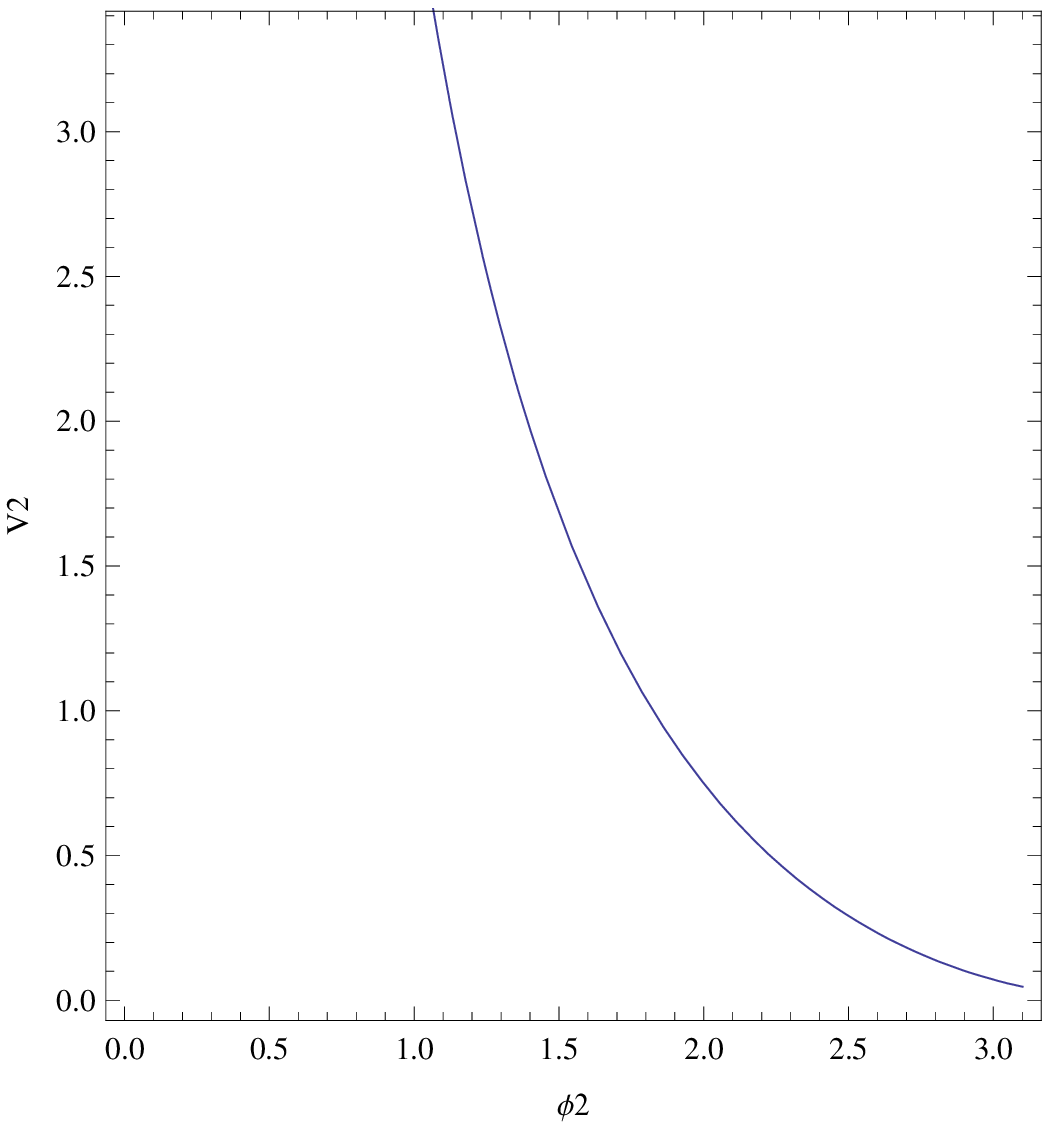}\\
\vspace{1mm} ~~~~~~~~~~~~Fig.13~~~~~~~~~~~~~~~~~~~~~~~~~~~~~~~~~~~~~~~~~~~~~~~~Fig.14\\

\vspace{7mm}

\includegraphics[height=1.8in]{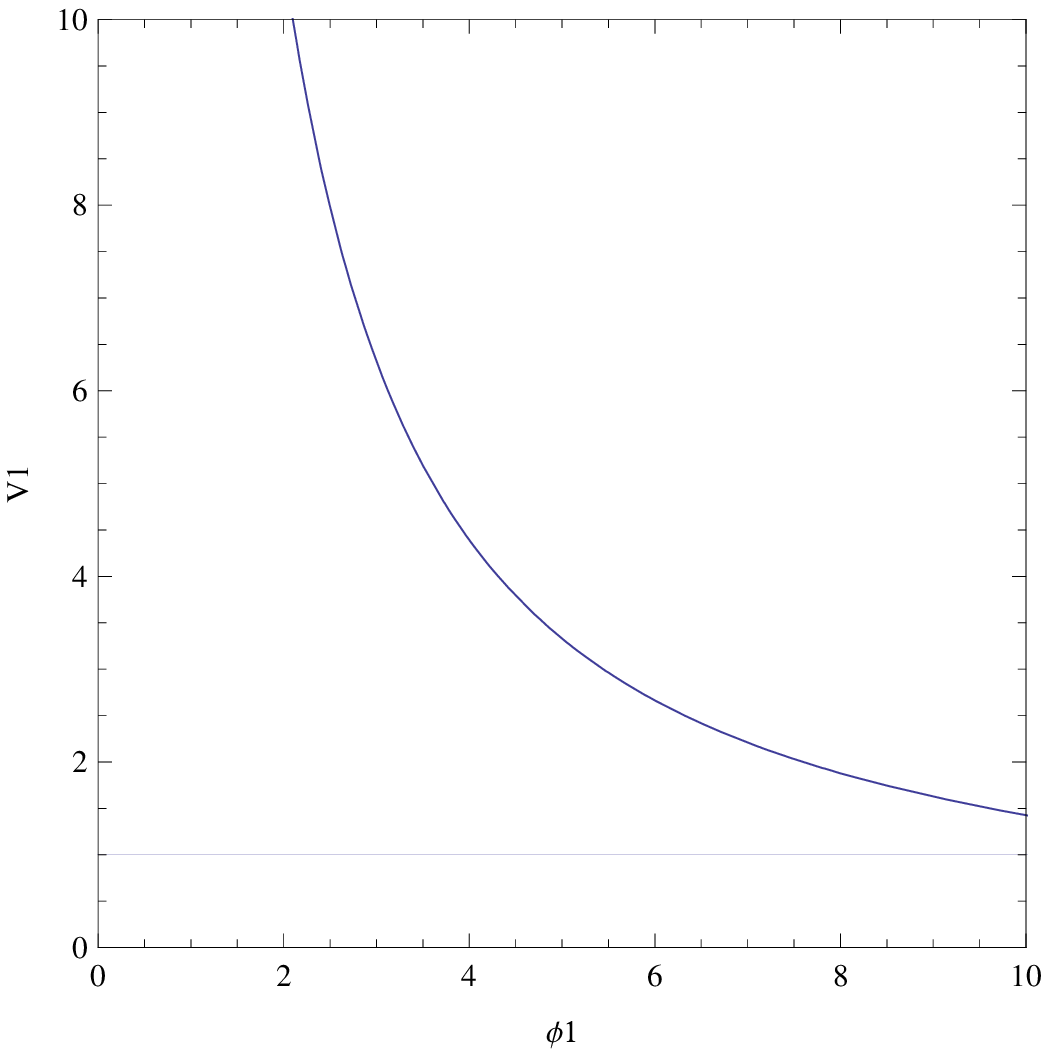}~~~~
\includegraphics[height=1.8in]{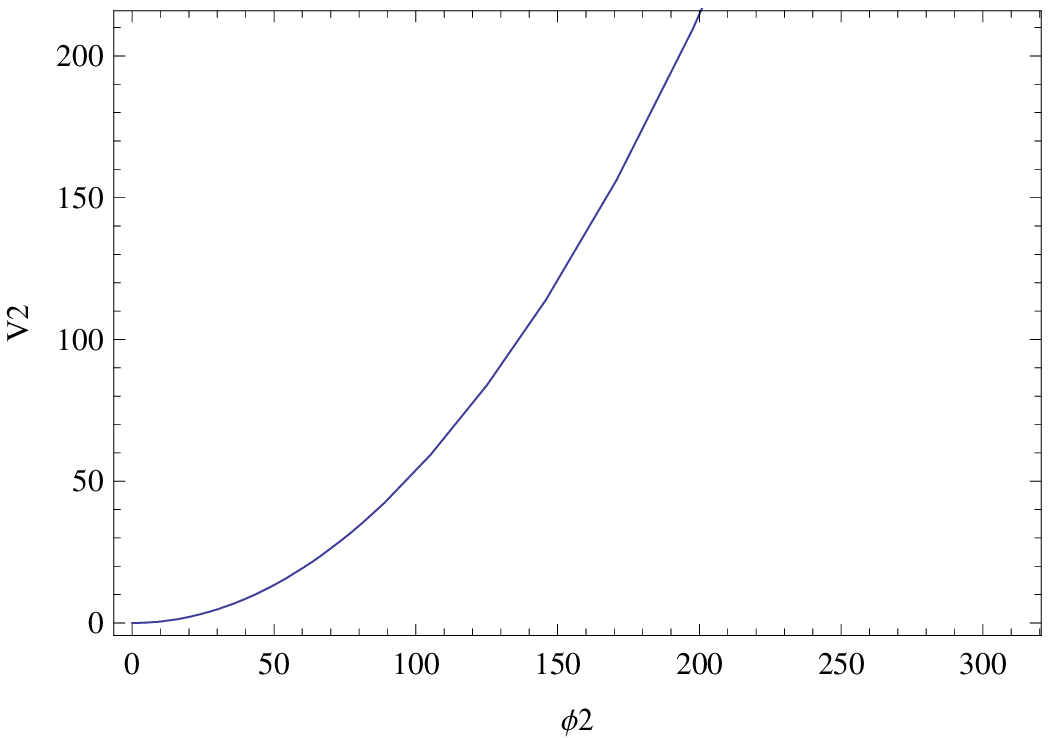}\\
\vspace{1mm} ~~~~~~~~~~~~~Fig.15~~~~~~~~~~~~~~~~~~~~~~~~~~~~~~~~~~~~~~~~~~~~~~~Fig.16\\

\vspace{7mm}

Figs. 13, 14 show the variations of $V_{1}$ with $\phi_{1}$,
$V_{2}$ with $\phi_{2}$ respectively for $\epsilon=+1, n=2,
\delta=0, m=c=c1=1$ and Figs. 15, 16 show the variations of
$V_{1}$ with $\phi_{1}$, $V_{2}$ with $\phi_{2}$ respectively for
$\epsilon=-1, n=2, \delta=0, m=c=c1=1$.\\

\vspace{6mm}

\end{figure}

{\bf Acknowledgement:}\\\\
One of the authors (UD) is thankful to UGC, Govt. of India for
providing research project grant (No. 32-157/2006(SR)).\\\\

{\bf References:}\\
\\
$[1]$  N. A. Bachall, J. P. Ostriker, S. Perlmutter and P. J.
Steinhardt, {\it Science} {\bf 284} 1481 (1999).\\
$[2]$ S. J. Perlmutter et al, {\it Astrophys. J.} {\bf 517} 565
(1999).\\
$[3]$ V. Sahni and A. A. Starobinsky, {\it Int. J. Mod. Phys. A}
{\bf 9} 373 (2000).\\
$[4]$ P. J. E. Peebles and B. Ratra, {\it Rev. Mod. Phys.} {\bf
75} 559 (2003).\\
$[5]$ T. Padmanabhan, {\it Phys. Rept.} {\bf 380} 235 (2003).\\
$[6]$ E. J. Copeland, M. Sami, S. Tsujikawa, {\it Int. J. Mod.
Phys. D} {\bf  15} 1753 (2006).\\
$[7]$ I. Maor and R. Brustein, {\it Phys. Rev. D} {\bf 67} 103508
(2003); V. H. Cardenas and S. D. Campo, {\it Phys. Rev. D} {\bf
69} 083508 (2004); P.G. Ferreira and M. Joyce, {\it Phys.
Rev. D} {\bf 58} 023503 (1998).\\
$[8]$ A. Melchiorri, L. Mersini, C. J. Odmann and M. Trodden, {\it Phys. Rev. D} {\bf 68} 043509 (2003).\\
$[9]$ A. Feinstein, {\it Phys. Rev. D} {\bf 66} 063511 (2002).\\
$[10]$ M. Sami, {\it Mod. Phys. Lett. A} {\bf 18} 691 (2003).\\
$[11]$ A. Sen, {\it JHEP} {\bf 0204} 048 (2002).\\
$[12]$ A. Sen, {\it JHEP} {\bf 0207} 065 (2002).\\
$[13]$ G. W. Gibbons, {\it Phys. Lett. B} {\bf 537} 1 (2002).\\
$[14]$ M. Sami, P. Chingangbam and T. Qureshi, {\it Phys. Rev. D}
{\bf 66} 043530 (2002).\\
$[15]$ M. Fairbairn and M.H.G. Tytgat, {\it Phys. Lett. B} {\bf 546}
1 (2002).\\
$[16]$ T. Padmanabhan, {\it Phys. Rev. D} {\bf 66} 021301 (2002).\\
$[17]$ J. S. Bagla, H. K. Jassal and T. Padmanabhan, {\it Phys.
Rev. D} {\bf 67} 063504 (2003); E. J. Copeland, M. R. Garousi, M.
Sami and S. Tsujikawa, {\it Phys. Rev D} {\bf 71} 043003 (2005);
G. Calcagni and A. R. Liddle, {\it Phys. Rev. D} {\bf 74} 043528 (2006).\\
$[18]$ M. Sami, P. Chingangbam and T. Qureshi, {\it Pramana}
{\bf 62} 765 (2004).\\
$[19]$ L. Parker and A. Raval, {\it Phys. Rev. D} {\bf 60} 063512
(1999); D. Polarski and A. A. Starobinsky, {\it Phys. Rev. Lett.}
{\bf 85} 2236 (2000); R. R. Caldwell, {\it Phys. Lett. B} {\bf 545} 23 (2002).\\
$[20]$ M. S. Berger, H. Shojaei, {\it Phys. Rev. D} {\bf 74}
043530 (2006).\\
$[21]$ R. Herrera, D. Pavon, W. Zimdahl, {\it Gen. Rel. Grav.}
{\bf 36} 2161 (2004); R. -G. Cai and A. Wang, {\it JCAP} {\bf
0503} 002 (2005); Z.-K. Guo, R.-G. Cai and Y.-Z. Zhang, {\it
JCAP} {\bf 0505} 002 (2005); T. Gonzalez and I. Quiros, {\it gr-qc}/0707.2089.\\
$[22]$ S. Chattopadhyay, U. Debnath and G. Chattopadhyay, {\it
Astrophys. Space Sci.} {\bf 314} 41 (2008).\\

\end{document}